\documentclass[aps,prb, twocolumn,superscriptaddress,groupedaddress,nofootinbib]{revtex4-1} 

\setcitestyle{numbers,square}

\usepackage{chngcntr}
\usepackage{array,multirow,graphicx}
\usepackage{dcolumn}  
\usepackage{bm}       
\usepackage{amssymb,amsmath}   
\usepackage[symbol]{footmisc} 
\usepackage{mathrsfs}
\usepackage{verbatim}
\usepackage{upgreek}
\usepackage[utf8]{inputenc}
\usepackage{pbox}
\usepackage{float}
\usepackage[caption = false]{subfig}

\begin{document}

\title{The $\phi^4$ theory Hamiltonian for fluids:\\
Application to the surface tension near the critical point}

\author{A.R. Dzhanoev}
\email{janoev@polly.phys.msu.ru}
\affiliation{M.V. Lomonosov Moscow State University, Faculty of Physics, Moscow, 119992, Russia}

\author{I.M. Sokolov}
\address{Institut f\"ur Physik, Humboldt-Universit\"at zu Berlin, Newtonstra{\ss}e 15, D-12489 Berlin, Germany}
\affiliation{IRIS Adlershof, Zum Gro{\ss}en Windkanal 6, 12489 Berlin, 
Germany}

\date{\today}

\begin{abstract}
We show that the surface tension of fluid near the critical point may be correctly described by taking into consideration the microscopic structure of 
the system using a $\phi^4$ field theory. 
We revise the theory of the surface tension near criticality to take into account a microscopic structure of the fluid. Focusing on the 
case of the Lennard-Jones fluid, we express the surface tension in terms of the compressibility of the reference 
hard-core system and its derivatives with respect to density. We demonstrate that the obtained {\it analytical} microscopic expression for the 
surface tension near the critical point is in a good agreement with numerical experiments, which emphasizes the impact of microscopic structure on the critical behavior of the surface tension in fluids. Our analysis provides a basis for studying the surface tension in small-volume systems important in many technological applications.
\end{abstract}

\maketitle
\section{INTRODUCTION}
The behavior of the surface tension (ST) has been a subject of continuous experimental and theoretical attention for many decades 
\cite{FiskWidom, Widom1972, Widom, Evans, Rayermann, Puibasset, Shishulin, NanoWodka, Pressing, BrezinFeng, Jug, Moldover, Mon, Muenster, Brackbill, 
Mecke, Potoff, Brilliantov2002, Caillol2010, 
Brezin2010, Ghoufi, Goujon, Popinet, Hernandez-Munoz, Eltsov, Lo, Kaurin}, with a wide variety of applications including the membranes in living cells \cite{Rayermann}, molecular films of nano-phase separated 
mixtures \cite{NanoWodka}, the nucleation \cite{Widom}, and the stability of spin-glass phases \cite{Brezin2010}. On the nano-scale, among other things, the attention is also focused on the phase 
transitions in liquid solutions within fractal nanoscale pores, where the behavior of the surface tension can be the dominant mechanism for the 
stratification \cite{Puibasset, Shishulin}. To date, theoretical studies of the ST usually involve density functional theory, which often gives a reasonable estimate of the ST \cite{Widom1972, Widom, Evans}. 
Moreover, the density functional theory can not be applied beyond the 
mean field (MF) level to study the critical region, where critical fluctuations should be considered. In this paper, within the formalism of the $\phi^4$ 
field theory we show that the surface tension of fluid near the critical domain may be properly described by taking into account the microscopic 
structure of the system. The derivation of the ST of fluid from the microscopic structure is of fundamental importance for basic science as well as for applications 
\cite{Brezin2010, NanoWodka, Rayermann, Puibasset, Shishulin} ranging from biophysics to technological processes. The detailed understanding 
and appropriate interpretation of the ST on the microscopic level is strongly desired because such knowledge is, for instance, important for wetting 
processes and capillary effect in small-volume systems.

The surface tension near criticality in the thermodynamic limit for a $l^d$ cubic lattice system vanishes as 
$\gamma (t, l\rightarrow \infty) = \gamma_0t^{\upmu}$, where $t \equiv (T_c - T)/T_c$, and $t > 0$, $\upmu$ is the interfacial tension critical exponent, and $\gamma_0$ 
is the amplitude. Below the critical temperature $T_c$ of the fluid there is the surface tension $\gamma$ that appears in an interface between two pure coexisting  phases of the system, defined as the free energy of inhomogeneity per unit area on the interface \cite{Widom1972, Widom}. 
The vicinity of the critical point can be described by a continuum model with the Hamiltonian of the $\phi^4$ field theory written in terms of a one-component order parameter $\phi \equiv \phi(x)$ in an arbitrary external field $h \equiv h(x)$:
\begin{eqnarray}
\begin{aligned}
\label{PHI4}
&\beta H = \int d^d x \left[ \frac{(\nabla \phi)^2}{2} - \tau_0 \frac{\phi^2}{2!} +g_0\frac{\phi^4}{4!} - h \phi \right],
\end{aligned}
\end{eqnarray}
where $\beta$ is the reciprocal of Boltzmann's constant times the 
temperature $T$, $d$ is dimensionality, and the coefficients $\tau_0$, $g_0$ are functions of $T$. The order parameter $\phi$ is of different physical 
nature in different systems and may be: density, composition of the fluid mixture, 
magnetization, etc. The universality of the transition at the critical point allows to consider a mapping of the system under 
consideration on a continuous scalar real $\phi^4$ field model \cite{ Zinn-Justin, Brezin}. The field-theoretical methods, elaborated especially for the 
$\phi^4$ field theory may be then applied to describe the critical fluctuations and derive an analytical expression for the surface tension as a function of parameters of the Hamiltonian (\ref{PHI4}). In the literature, for the $\phi^4$ field theory model, it is common to consider the 
coefficients of the Hamiltonian without reference to their connection with the microscopic structure of the system \cite{Zinn-Justin, Brezin, BrezinZinn, Parola, Pelissetto}. However, a series of measurements of the ST covering a wide range of temperatures 
\cite{Mecke, Potoff} shows that such an approach fails to give a satisfactory explanation of the experimental MD and MC data at critical domain. The analytical formula of the ST 
developed by Br\'ezin and Feng \cite{BrezinFeng} by means of 
Wilson's $\epsilon$-expansion \cite{WilsonKogut}, where $\epsilon = 4 - d$, generally shows a better agreement with experiments for $\epsilon \rightarrow 1$, but still 
there is some discrepancy with the data \cite{Moldover}. M\"unster \cite{Muenster} suggested an alternative analysis in the framework of quantum field 
theory, and derived the ST for $d=3$. However, in all these approaches the choice of the coefficients of the Hamiltonian is to a great extent arbitrary. 
One of the known attempts to include the microscopic structure into analytical consideration of the ST for fluids in the case of the hard spheres 
reference system was made in \cite{Brilliantov2002}. But, the announced result was never published to the best of our 
knowledge. It uses the approach derived by Hubbard and Schofield \cite{Hubbard} where 
the effective Landau-Ginzburg-Wilson (LGW) Hamiltonian for fluids was derived by an exact mapping, based on the transformation of variables. 
Still, these studies of the ST and the fluid criticality by means of the LGW Hamiltonian approach were performed 
within the MF approximation where critical fluctuations are not considered. An alternative analysis with the inclusion of 
microscopic structure into the consideration of the ST based on the statistical field theory for a fluid was made by Caillol in \cite{Caillol2010}, where a 
qualitative description of a planar liquid-vapor interface was obtained.

In this paper, we study the surface tension of fluid near the critical point by considering the microscopic structure of the system using a 
$\phi^4$ field theory. For this, we express 
the coefficients of the $\phi^4$ theory Hamiltonian in terms of the compressibility of the reference system and its derivatives with respect to density.  We show that the derived analytical microscopic expressions for the ST allows to obtain this quantity for the given temperature, density, and 
interaction potential. Further, in order to verify our analytical results, we also derive the microscopic expression for the critical behavior of the ST obtained using the parameters of the LGW-Hamiltonian for fluids. Thus, at the MF level, we show that our analysis accurately reproduces dependencies for the microscopic expression of the ST obtained by a different approach \cite{Brilliantov2002}. We show that using the microscopic interpretation of parameters of the $\phi^4$ Hamiltonian we achieve the qualitatively 
and quantitatively good agreement with the numerical simulations of the ST for fluids near the critical point. 

\section{METHOD AND MODEL} 
The ST is an essential characteristic of material interfaces. Physically, liquid surfaces are in a state of tension because fluid molecules at the surface or near it experience unequal molecular forces of attraction. The ST leads to a microscopic localized ``surface force'' that acts on fluid elements at interfaces in both the normal and tangential directions. 

To highlight the impact of microscopic structure of fluid on the ST, we consider the one-component system of classical particles interacting via a pairwise potential $w(r)$ which is supposed to be short-range ($r^{d+2} w(r) \rightarrow 0$ in $d$ dimensions as $r\rightarrow \infty$), such that the system possesses a thermodynamic limit and a 
liquid-vapor critical point. In addition, we assume that the interaction potential $w(r)$ may be resolved into an attractive part $-v(r)\leqslant 0$ and the repulsive part $\varphi(r) \geqslant 0$ in such a way that $v(r)$ is bounded and has a positive Fourier transform $v_k > 0$ with the property 
$v_k = v_0 - v^{\prime\prime} k^2 + ...$, for $k$ small; $\varphi(r)$ should be as short-range as possible consistent with these conditions. 
In order to be more specific, we shall focus on the Lennard-Jones (LJ) pairwise interaction potential $w(r)=4  \varepsilon[(\sigma/r)^{12} - (\sigma/r)^6]$ in three dimensions ($d=3$). We use the Weeks-Chandler-Andersen (WCA) partition \cite{WCA} of the LJ potential into attractive and repulsive parts that gives for the attractive part: 
\begin{equation}
\label{LJP}
v(r)=
\begin{cases}
  \varepsilon, \hspace{9.5mm} r\leqslant2^{1/6}\sigma\\      
  -w(r), \hspace{2mm} r\geqslant2^{1/6}\sigma,
\end{cases}
\end{equation}
which is smooth in the core region, and for the repulsive part: $\varphi(r) = w(r) + v(r)$. This partition provides the best estimates for the thermodynamic functions in the WCA perturbation scheme \cite{WCA}. 
For the reference system with the only repulsive interactions we use the hard-sphere system with an 
appropriately chosen diameter \cite {Gubbins}: $D = \displaystyle\int_{0}^{\sigma} dr \left[ 1 - \exp(-\beta \varphi(r))\right]$ 
that gives the effective diameter of the hard-sphere system, corresponding to a repulsive potential $\varphi(r)$ vanishing at $r \geqslant \sigma$. 
Also, it reproduces the second virial coefficient for the repulsive part. Note that recently, the case when the reference system in addition to the hard-sphere repulsion includes also 
the short range attraction was considered by Trokhymchuk et al. \cite{Trokhym}. For the hard-core system \cite{Mecke, Holcomb}, one has the Carnahan-Starling equation of state, which for the reduced isothermal compressibility yields $z_0 = (1-\eta)^4 / (1 + 4\eta + 4\eta^2 - 4\eta^3 +\eta^4)$, 
where $\eta = \displaystyle\frac{1}{6}\pi \rho \hspace{.5mm} D^3$ - the packing fraction, and derivatives of $z_0$ with respect to density $\rho$ are 
defined as $z_1 \equiv \rho \partial z_0 / \partial \rho$, $z_2 \equiv \rho^2 \partial^{2}z_0/\partial \rho^2$. The compressibility $z_0$ can be related to 
the zero-$k$ value of the Fourier transform of the total correlation function $\tilde{h}(\mathbf{k})$
as $1+\rho\tilde{h}_2(0)=\beta^{-1}\rho\chi_R\equiv z_0$ with $\chi_R=\rho^{-1}(\partial \rho / \partial p_r)_{\beta}$, where $p_r$ is the pressure of the reference fluid. The function $\tilde{h}_2(0)$ may 
be expressed in terms of the zero-$k$ value of the Fourier transform of the direct correlation function $\tilde{c}(\mathbf{k})$ 
by means of the Ornstein-Zernike relation (see also Appendix A). The functions $\tilde{h}(\mathbf{k})$, and $\tilde{c}(\mathbf{k})$ are commonly 
used for describing the microscopic structure of a simple fluid \cite{Gubbins}.

To describe the surface tension $\gamma(\tau)$ close to the critical point in the one-loop approximation, we use an expression for 
$\gamma(\tau)$ derived by Br\'ezin and Feng \cite{BrezinFeng}
\begin{eqnarray}
\begin{aligned}
\label{STRenorm}
&\frac{\gamma(\tau)}{kT} = \\
&\displaystyle\frac{4\sqrt{2}}{g} \left[1 + \frac{\epsilon}{4}\  \left( 1 - \ln 2 - \frac{\pi\sqrt{3}}{9} \right) \right]\tau^{3/2-\epsilon/4},
\end{aligned}
\end{eqnarray}
which is written in terms of renormalized parameters $\tau$, $g$ of the $\phi^4$ Hamiltonian (\ref{PHI4}), and relate these parameters to the 
microscopic structure of the LJ-fluid. In the MF approximation we have \cite{BrezinFeng}
\begin{equation}
\label{STMF}
\frac{\gamma^{MF}}{kT} = \displaystyle\frac{4\sqrt{2}}{g}\tau^{3/2},
\end{equation}
which yields the van der Waals surface tension exponent $\upmu =3/2$. We notice that the renormalized parameters $\tau$, $g$ are associated with the parameters 
$\tau_0$, and $g_0$ through the relations: $\tau_0 Z=\tau Z_2$, and $g_0 Z^2=\Lambda^{\epsilon} g Z_1$ (see Appendix B). Here, the minimal subtraction renormalization scheme is used, in which $Z$'s are power series in $g$ with coefficients containing 
only multiple poles in $\epsilon$, but no finite part \cite{Zinn-Justin,BrezinFeng} (also see Eq.(24) in \cite{BrezinFeng} for the one-loop 
$Z$-counterterms). 

First, we derive the field theoretical Hamiltonian from the fluid-Hamiltonian (\ref{FluidHamn}) (see Appendix A). Then, in order to take into 
account the relationship between parameters of the $\phi^4$ Hamiltonian (\ref{PHI4}) and the microscopic structure of the LJ-fluid, we express 
these parameters in terms of the compressibility $z_0$ of the  reference system and its derivatives $z_i$ with respect to density. Thus, we 
arrive at the $\phi^4$ field theory Hamiltonian (\ref{PHI4}) with new generalized microscopic expressions for its coefficients that in three 
dimensions ($d=3$) are written as 
\begin{eqnarray}
\begin{aligned}
\label{transform0}
& \tau_0 =  \rho z_0(\rho_c\kappa)^{-1} + u_3^2 (2 g_0\kappa^3)^{-1} - (\beta v_0\rho_c \kappa)^{-1}\\ 
& g_0 = - (\rho/\rho_c) z_0 \kappa^{-2} \left[ z_1^2 + z_0(z_0 + 4z_1 + z_2)\right]\\
& h = \mu^{\prime} (v_0 \rho_c \kappa)^{-1} + \rho (\rho_c\kappa)^{-1} \\
& - \left[ \tau_0 - \frac{u_3^2}{6g_0 \kappa^3} \right]\frac{u_3}{g_0 \kappa^2}\\
& \kappa = \displaystyle\frac{3\rho_c^{-1/3}}{40\pi D} \left[ \lambda_{e}^2 (\beta \varepsilon_{e})^{-1} - \Delta \right],
\end{aligned}
\end{eqnarray}
where $\mu^{\prime} \equiv \mu -(\mu_R+v_0/2)$ with $\mu$ - the chemical potential of the fluid with the pairwise LJ potential $w(r)$, while $\mu_R$ is the chemical potential in the reference system, $u_3 \equiv -(\rho/\rho_c) z_0 (z_0 + z_1)$, $\rho_c$ is the critical density, $\Delta = 
4 \eta^2 (1 - \eta)^4 (16 - 11\eta +4 \eta^2) / (1 + 4\eta + 4\eta^2 - 4\eta^3 + \eta^4)^2$, and constants $\varepsilon_{e}$ and $\lambda_{e}$ 
characterize the effective depth and the effective width of the attractive part $v(r)$ of the LJ potential (\ref{LJP}): $\varepsilon_{e} = (4\pi D^3/3)^{-1}\displaystyle\int 
v(\mathbf{r}) d\mathbf{r} = (4\pi D^3/3)^{-1} v_0$, and 
$\lambda_{e}^2 = (3v_0 D^2/5)^{-1} \displaystyle\int v(\mathbf{r})r^2 d\mathbf{r} = (3v_0 D^2/5)^{-1} v^{\prime\prime}_0$. 
Thus, the compressibility and its derivatives with respect to the density in (\ref{transform0}), in its turn, are  related with the intermolecular interaction 
potential. For details we refer the reader to Appendix A. 

At the critical point the two phases merge into one homogeneous phase. The average density of our system is $\overline{\rho} =(\rho_L + \rho_G) / 2 \simeq \rho_c$, and thus one can use $\rho_c$ as the reference density (see e.g. Fig.\ref{PhasDiag}, Top). 
Specifically, following the analysis near the critical point \cite{Zinn-Justin}, one can write for $\tau_0$: $\tau_0 \simeq \tau_{0c} \equiv  [\rho z_0(\rho_c\kappa)^{-1} + u_3^2 (2 g_0\kappa^3)^{-1}]_c - (\beta v_0\rho_c \kappa_c)^{-1}$
as it follows from (\ref{transform0}). If then we use the condition for the critical point $\tau_0(\beta_c,\rho_c)=0$, 
we obtain $\tau_{0c} = (\beta_c v_0\kappa_c\rho_c)^{-1} -  (\beta v_0\kappa_c\rho_c)^{-1} = \kappa_c^{-1}\alpha t$. Finally, we obtain a new generalization of the commonly used expression (\ref{STRenorm}) for the surface tension in the one-loop approximation as 
\begin{eqnarray}
\begin{aligned}
\label{STRG}
&\frac{\gamma(\tau)}{kT} = \\
&\displaystyle\frac{4\sqrt{2}}{g_{0c}} \left[1 + \frac{\epsilon}{4}\  \left( 1 - \ln 2 - \frac{\pi\sqrt{3}}{9} \right) \right](\kappa_c^{-1}\Lambda^2 \alpha t)^{3/2-\epsilon/4},
\end{aligned}
\end{eqnarray}
and in the MF approximation from (\ref{STMF}) we have
\begin{equation}
\label{STMFRG}
\frac{\gamma^{MF}}{kT} = \displaystyle\frac{4\sqrt{2}}{g_{0c}}(\kappa_c^{-1} \alpha t)^{3/2},
\end{equation}
where $\alpha = (\beta_c v_0 \rho_c)^{-1} = (8\beta_c \eta_c \varepsilon_e)^{-1}$, and the coefficients $g_{0c}$ and $\kappa_c$ are to be calculated at $\rho=\rho_c$, $T=T_c$. At critical temperatures, in the field-theoretical description the parameter $\Lambda$ is an arbitrary inverse length 
scale \cite {Zinn-Justin} (see also Appendix B). However, in our approach this parameter has a certain physical meaning and measured in 
$\rho_c^{1/3}$ units (in three dimensions) - the inverse mean molecular distance near the critical point.

We note that from (\ref{STRG}) and (\ref{STMFRG}) we can obtain expressions for the ST in terms of the coefficients of the LGW Hamiltonian 
$\mathscr{H}$ (\ref{LGW}) in the MF and one loop approximations (see Appendix B).  In the critical domain, repeating the analysis done for the coefficients of the $\phi^4$ Hamiltonian, we obtain: 
$a=- (\beta_c v_0 \rho_c)^{-1}t \equiv- \alpha t$, and the coefficients $b_{c}$ and $\kappa_c$ are calculated at the critical point. 
In particular, from the equation (\ref{STMFRG}) for the surface tension in the MF approximation, for $d=3$ we have
\begin{equation}
\label{MFGammaCR}
\frac{\gamma^{MF}}{k T} = \frac{4\sqrt{2}}{b_c} (\kappa_c^{1/3}\alpha t)^{3/2}.
\end{equation}
The expression (\ref{MFGammaCR}) reproduces the analytical microscopic expression for the MF surface tension derived in \cite{Brilliantov2002}. 
So, this verifies our approach to the critical behavior of the surface tension at the microscopic level.

\section{RESULTS} 
Now, we apply the analytical expressions for the surface tension (\ref{STRG}) and (\ref{STMFRG}) with the microscopic parameters defined 
by (\ref{transform0}), to the case of space dimension $d=3$, by using the extrapolation $\epsilon \rightarrow 1$ in the one-loop case, 
and compare the results with the available data of numerical experiments for the Lennard-Jones fluid. 

\begin{figure}[ht!]
\centering
\includegraphics[scale=1.0]{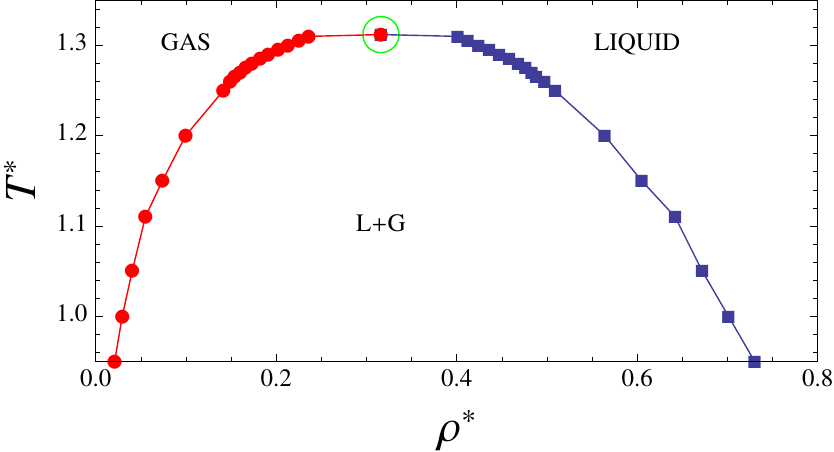}
\\ \vspace{.5cm}
\includegraphics[scale=1.0]{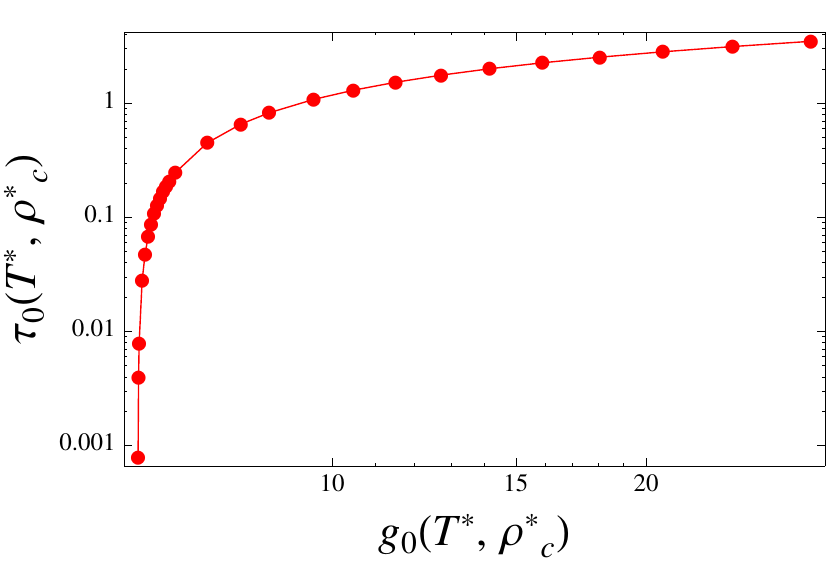}
\caption{(Color online) {\it Top}: The liquid-vapor phase diagram in the case of the LJ-fluid derived in \cite{Potoff}. In the green circle the critical point: 
$( \rho^*_c, T^*_c ) = (0.316, 1.312)$ is marked. Here, we consider the variables that reduced as follows: the temperature $T^*=kT/\varepsilon$, 
the density $\rho^*=\rho \sigma^3$, and the surface tension $\gamma^* = \gamma\sigma^2/\varepsilon$, where $\sigma$, $\varepsilon$ are parameters 
of the LJ potential; {\it Bottom}: The parameters $(g, \tau)$ of the $\phi^4$ Hamiltonian calculated at the reduced density $\rho^*_c$ and reduced 
temperatures $T^*$ given in Fig. \ref{PhasDiag}({\it top}). Here $\alpha(\rho^*_c,T^*_c) = 0,2628$, $\kappa(\rho^*_c,T^*_c) = 0,0506$. The lines serve as a guide to the 
eye.}
\label{PhasDiag}
\end{figure}

In Fig.\ref{PhasDiag} (Bottom) we show the plane of parameters ($g,\tau$) of the $\phi^4$ Hamiltonian calculated for 
the reduced temperatures and the reduced densities obtained in \cite{Potoff} and shown in diagram Fig.\ref{PhasDiag} (Top). 

\begin{figure}[ht!]
	\centering
	\includegraphics[scale=1.0]{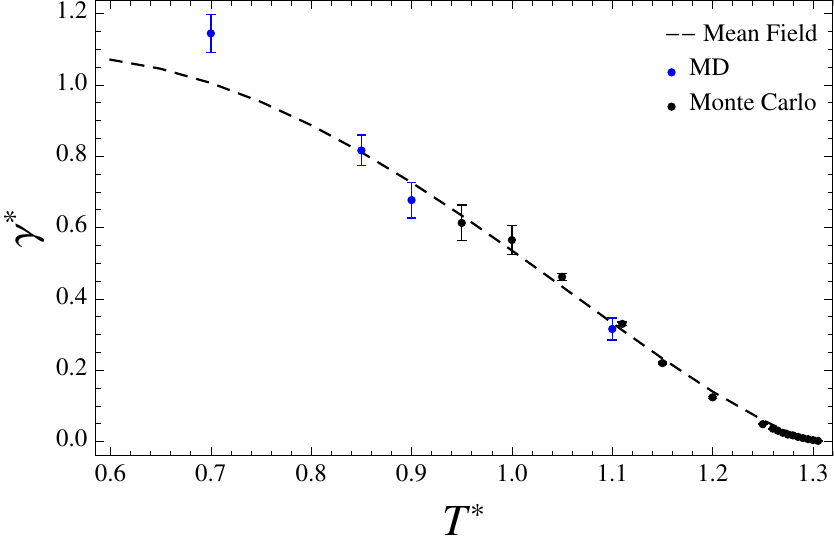}
	\\ \vspace{.5cm}
	\includegraphics[scale=1.035]{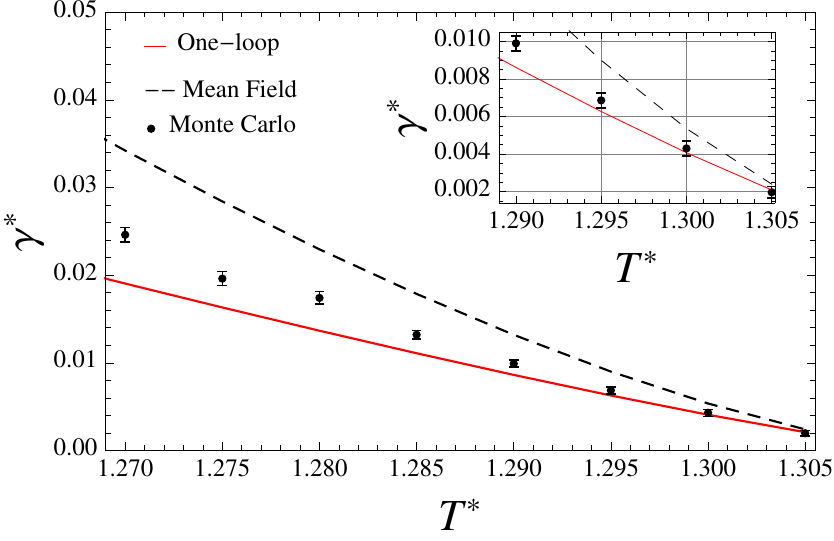}
	\caption{(Color online) {\it Top}: The reduced surface tension $\gamma^*$ as a function of the reduced temperature 
		$T^*$ for LJ-fluid. The dashed curve - the MF theory as it follows from (\ref{STMFRG}). Points - numerical data for the MD (blue) and MC 
		(black) simulation of LJ-fluid \cite{Mecke, Potoff}; 
		{\it Bottom}: The reduced surface tension for the LJ fluid versus $T^*$ near the critical point. The dashed curve -- MF theory, the solid curve -- the one-loop correction, Eq.(\ref{STRG}). Points - numerical data for the MC simulations of LJ-fluid \cite{Potoff}. {\it Inset:}  zoom for temperatures close to 
		$T_c^*=1.312$ emphasizing the good agreement of the one-loop result (\ref{STRG}) for the surface tension with the MC experiment. 
		Critical parameters for the LJ-fluid are taken from \cite{Potoff}. $\sigma$, $\varepsilon$ are parameters of the LJ potential \cite{Potoff}.}
	\label{GamBrez}
\end{figure} 
In Fig.\ref{GamBrez} we compare the estimates of the MF theory for the ST and its one-loop correction with experimental results. 
As follows from Fig.\ref{GamBrez} (Top) our theoretical prediction within MF approach is close to results of the numerical MC and MD experiments, 
except for the very close vicinity of the critical point, where the mean field theory loses its accuracy. Note that, the Fig.\ref{GamBrez} (Top) 
reproduces the analogous MF dependence of the ST from the temperature obtained in \cite{Brilliantov2002}. 
In Fig.\ref{GamBrez} (Bottom) we see that in the very close vicinity of the critical point our results for the ST in the one-loop approximation converge better to experimental data than in the MF case.

\section{CONCLUSION} 
In the present work, in the framework of the $\phi^4$ field 
theory model, we described the critical behavior of the surface tension derived from the microscopic structure of fluid. We illustrated that for the LJ-fluid near the critical point, the {\it analytical} expression of the surface tension through microscopic parameters of the $\phi^4$ Hamiltonian, allows to obtain a good agreement with MD and MC simulations. 

Finally, our results demonstrate the importance of including the microscopic structure into the description of the surface tension in fluids, 
which would be helpful for better understanding of the experimental studies of the critical phenomena. The calculations for the surface tension 
we presented in this paper in the framework of the $\phi^4$ field theory are general, and can be readily extended to the microscopic models of other 
physical quantities near the critical point. 

\section{ACKNOWLEDGMENTS} 
Authors acknowledge helpful discussions with N.V.Brilliantov who also has turned our attention to this problem.
\begin{appendix}
\subsection*{APPENDIX A}
\setcounter{equation}{0}
\renewcommand{\theequation}{A\thesection.\arabic{equation}}

In 1972 Hubbard and Schofield showed that if the pairwise interaction potential $w(r)$ can be split into repulsive $\phi(r)$ and attractive 
part $-v(r)$, then the fluid Hamiltonian
\begin{equation}
\label{FluidHamn}
H^{\text{fl}} = \sum_{i<j} w(r_{ij})=\sum_{i<j} \phi(\mathbf{r}_{ij}) - \sum_{i<j} v(\mathbf{r}_{ij}),
\end{equation}
can be mapped, by means of the Hubbard-Stratonovich transformation \cite{Hubbard}, onto the effective field theoretical Hamiltonian
\begin{eqnarray}
\begin{aligned}
\label{Hamn}
&\beta \mathcal{H} = \\ 
&-\tilde{h} V^{1/2}\phi_0 + \displaystyle\sum_{n=2}^{\infty} V^{1-n/2}\sum_{\mathbf{k}_1,...,\mathbf{k}_n} 
\tilde{u}_n\phi_{\mathbf{k}_1} \dotsm \phi_{\mathbf{k}_n},
\end{aligned}
\end{eqnarray}
where $V = L^3$ is the volume of the system, and we imply the summation over the set of $\mathbf{k}$:
 $k_s = \displaystyle \frac{2\pi}{L}m_s$ with $s = x, y, z$, and $m_s = 0, \pm1,\pm2, ...$; also we assume the thermodynamic limit 
 $L\rightarrow \infty$. The coefficients of the Landau-Ginzburg-Wilson (LGW) Hamiltonian $\mathcal{H}$ read as
$\tilde{h} = \mu^\prime v_0^{-1} + \rho$, and 
\begin{eqnarray}
\begin{aligned}
\label{Koff}
& \tilde{u}(\mathbf{k}_1,\mathbf{k}_2) = \displaystyle\frac{1}{2!} \delta_{\mathbf{k}_1+\mathbf{k}_2,0}\left [ \beta^{-1} v_{\mathbf{k}_1}^{-1} - \left< n_{\mathbf{k}_1} n_{\mathbf{k}_2} \right>_{cR} \right ], \\
& \tilde{u}_n(\mathbf{k}_1,\dots\mathbf{k}_n) =  
 - \displaystyle\frac{V^{n/2-1}}{n!} \left< n_{\mathbf{k}_1} \dotsm n_{\mathbf{k}_n} \right>_{cR}, \text{for } n \geq 3.
\end{aligned}
\end{eqnarray}
Here, the Fourier components of the density $n_{\mathbf{k}} = \displaystyle V^{-1/2}\sum^N_{j=1}e^{-i\mathbf{k}\mathbf{r}_j}$, with $N$ - the number of particles, and the 
Fourier transform of the attractive potential $v_{\mathbf{k}} = \displaystyle \int v(\mathbf{r}) e^{-i\mathbf{k}\mathbf{r}} d\mathbf{r}$, 
$\rho = V^{-1/2}\left< n_0 \right>_{cR} = N/V$ is the fluid density, and $\left< \dots \right>_{cR}$ denotes the cumulant average calculated 
in the homogenous reference system. The cumulant average of a product excludes all products of cumulant averages of all subsets. 
The explicit form of the first few cumulants is \cite{Kubo}
\begin{eqnarray}
\begin{aligned}
\label{Cumul2Aver}
& \left< A_1 \right>_{c} = \left< A_1 \right>,\\
&  \left< A_1 A_2 \right>_{c} = \left<  A_1 A_2\right> - \left< A_1 \right> \left< A_2 \right>, \\
&  \left< A_1A_2A_3 \right>_c = \left<  A_1A_2A_3\right> - \left<  A_1A_2\right>_c \left< A_3 \right> \\ 
&  - \left<  A_2A_3\right>_c \left< A_1 \right> -  \left<  A_1A_3\right>_c \left< A_2 \right>  - \left< A_1 \right> \left< A_2 \right> \left< A_3 \right>,
\end{aligned}
\end{eqnarray}
while the general formula for calculating cumulants in terms of averages has been obtained in \cite{Meeron}. As it follows 
from Eq. (\ref{Koff}) the coefficients of $\mathcal{H}$ depend on the correlation function of the reference fluid with the repulsive 
interactions only. 

The next step in developing the fluid LGW Hamiltonian (\ref{Hamn}) was made in \cite{Brilliantov1998}, where by using Eqs. (\ref{Cumul2Aver}) and definitions 
of the particle correlation functions of fluid \cite{Gubbins}, the cumulant averages $ \left< n_{\mathbf{k}_1} \dotsm n_{\mathbf{k}_n} 
\right>_{cR}$ and thus the coefficients $\tilde{u}_n(\mathbf{k}_1,\dots\mathbf{k}_n)$ were expressed in terms of the Fourier transforms of the particle correlation 
functions of the fluid with the 
short range interaction. Also, in \cite{Zinn-Justin}, \cite{Brilliantov1998} it was shown that the effective Hamiltonian (\ref{Hamn}), after using $\rho_c^{-1/3}$ (for $d=3$) as a scaling factor for the length, can be transformed into the conventional form
\begin{eqnarray}
\begin{aligned}
\label{LGW}
&\beta \mathscr{H} = \int d\mathbf{r} \left[\kappa(\nabla \phi)^2/2 + \frac{a}{2!}\phi^2(\mathbf{r}) + \frac{b}{4!}\phi^4(\mathbf{r}) \right.\\
& - \left.  h(\mathbf{r})\phi(\mathbf{r}) \right],
\end{aligned}
\end{eqnarray}
which is used to describe the behavior of many systems with short range interactions, in the critical domain \cite{Zinn-Justin}.
The microscopic expressions for the parameters of the Hamiltonian $ \mathscr{H}$ were obtained in \cite{Brilliantov1998}.
Finally, after rescaling the field $ \kappa^{1/2} \phi 
\longmapsto \phi$, we arrive at the $\phi^4$ field theory Hamiltonian (\ref{PHI4}) with microscopic expressions for its 
coefficients (\ref{transform0}). Further, we study the $\phi^4$ Hamiltonian (\ref{PHI4}) with 
the coefficients (\ref{transform0}) by the field theoretical methods.
\end{appendix}

\begin{appendix}
\subsection*{APPENDIX B}
\setcounter{equation}{0}
\renewcommand{\theequation}{B\thesection.\arabic{equation}}

We will consider in some detail the epsilon expansion introduced by Wilson \cite{WilsonKogut} and 
later developed by many authors \cite{Zinn-Justin, Brezin, BrezinZinn, Parola, Pelissetto}. We demonstrate how it is applied to the analysis of 
the behavior of the critical surface tension using the $\phi^4$ theory. Here, we follow an analysis by Br\'ezin and Feng from \cite{BrezinFeng}. For the definiteness, we consider a sample contained in a vertical cylinder of height  $L$ and cross-sectional area $A$; if the spins point down 
in the $z = -L/2$ plane and up in the $z = L/2$ plane, an interface appears between two pure phases of opposite magnetization. 

The partition function for the system with the $\phi^4$ Hamiltonian (\ref{PHI4}):
\begin{equation}
\label{PartFunc}
\exp W[h] = \int D \phi e^{-\beta H},
\end{equation}
and the Legendre transform $\Gamma [\phi]$ of $W[\phi]$ defined by 
\begin{equation}
\begin{aligned}
\label{Legandre}
& \phi = \frac{\delta W}{\delta h},\\
& \Gamma [\phi] = -W[h] + \int d^dx h(x) \phi (x).
\end{aligned}
\end{equation}
In zero field the free energy is given by
\begin{equation}
\label{ZeroField}
\beta F = \Gamma[\phi^c],
\end{equation}
here $\phi^c(x)$ is the solution of 
\begin{equation}
\label{Solut}
h(x) \equiv \frac{\delta \Gamma}{\delta\phi(x)} =0.
\end{equation}
The Eq. (\ref{Solut}) must be supplemented by boundary conditions. For up-up or down-down boundary conditions at $z = -L/2$ and $z = L/2$, 
respectively, $\phi^c(x)$ is uniform and equal to the spontaneous magnetization $M_0$. For down-up boundary conditions, 
$\phi^c(z,\vec{x}_{||})$ is independent of $\vec{x}_{||}$ but vary with $z$ between $-M_0$ and $M_0$.

In order to go beyond the mean-field theory, we use the renormalized perturbation theory \cite{Zinn-Justin}, \cite{BrezinZinn}. The theory with Hamiltonian (\ref{PHI4}) is {\it regularized} by considering its extension in dimension $d = 4 - \epsilon$. 
Divergences are removed by expressing all bare couplings in terms of renormalized ones. Thus, in terms of 
renormalized parameters we have \cite{Zinn-Justin}
\begin{eqnarray}
\begin{aligned}
\label{RenormHamn}
& \beta H = \\
&\displaystyle\int d^d x \left[ \displaystyle\frac{Z(\nabla \varphi)^2}{2} - \tau\displaystyle\frac{Z_2 \varphi^2}{2} + \Lambda^{\epsilon}g\displaystyle\frac{Z_1 \varphi^4}{4!} \right],
\end{aligned}
\end{eqnarray}
where $\Lambda$ is equal to an arbitrary inverse length scale, so that $g$ is dimensionless; $\tau$, proportional to $\Lambda^2$, is a linear 
measure of the temperature. 

First, we consider a simpler model (mean-field) in which the lowest approximation, in which we neglect all fluctuations of the order parameter 
around its most likely value, reduces to the Landau theory. Thus, at the MF level ($\epsilon = 0$) for the free energy we have 
\cite{Zinn-Justin}
\begin{equation}
\label{MeanField}
\Gamma^{MF}[\varphi] = \displaystyle\int d^4 x \left[ \frac{(\nabla \varphi)^2}{2} - \tau\frac{\varphi^2}{2} + g\frac{\varphi^4}{4!} \right].
\end{equation}

The renormalized critical free-energy (at one-loop order) is \cite{Zinn-Justin}
\begin{eqnarray}
\begin{aligned}
\label{OneLoopG}
& \Gamma^{(1)}[\varphi] = \\
&\displaystyle\int d^d x \left[ \displaystyle\frac{Z(\nabla \varphi)^2}{2} - \tau\displaystyle\frac{Z_2 \varphi^2}{2} +
\Lambda^{\epsilon}g\displaystyle\frac{Z_1 \varphi^4}{4!} \right] \\
& + \frac{1}{2}\text{Tr} \ln K,
\end{aligned}
\end{eqnarray}
with $K = \delta(x-y)\left[ -Z \nabla^2 - \tau Z_2 +  \displaystyle\frac{1}{2} \Lambda^{\epsilon}gZ_1\varphi^2 \right]$.\\

There are several physical observables that might be drawn from (\ref{OneLoopG}); they would all lead to a similar analysis of the critical 
behavior. Here, we consider the critical behavior of the surface tension as it was done in \cite{BrezinFeng}.
Below $T_c$ there is the surface tension $\sigma$ that appears in an interface between two pure phases of opposite magnetization, defined 
in terms of the free energy $F$ per unit area $A$ as
\begin{equation}
\label{ST}
\sigma = (F_{\uparrow \downarrow} - F_{\uparrow\uparrow}) / A,
\end{equation}
where the arrows specify the boundary conditions in the planes $z =  L/2$ and $z = - L/2$ correspondingly. 
Noting that \cite{BrezinFeng}
\begin{eqnarray}
\begin{aligned}
\label{STGamma}
& \Gamma_{\uparrow \downarrow}^{MF} - \Gamma_{\uparrow\uparrow}^{MF} = \\
& A \displaystyle \int_{-\infty}^{+\infty} dz \left [ \frac{1}{2} \left( \frac{d \varphi_c}{dz} \right)^2 + \frac{1}{4!}g[\varphi_c^2 \right.\\
& \left.- (M^{MF})^2]^2 \right ],
\end{aligned}
\end{eqnarray}
in the MF approximation we obtain the expression (\ref{STMF}). Similarly, using the Eq. (\ref{OneLoopG}) the one-loop corrections to the MF expression (\ref{STMF})(the first order in $\epsilon$) for the surface tension, following Br\'ezin and Feng \cite{BrezinFeng}, gives the relation (\ref{STRenorm}). 
\end{appendix}\\


\begin{thebibliography}{99}

\bibitem{FiskWidom}
S.~Fisk, and B.~Widom, 
J. Chem. Phys. {\bf 50}, 3219 (1969).

\bibitem{Widom1972} 
B.~Widom, in {\it Phase Transitions and Critical Phenomena}, edited by C.~Domb, and M. S.~Green (Academic Press, London, UK, 1972),  {\bf 2}, p. 79.

\bibitem{Widom}
J. S.~Rowlinson, and B.~Widom,
{\it Molecular Theory of Capillarity} (Clarendon Press, Oxford, UK, 1982). 


\bibitem{Evans}
R.~Evans, in {\it Fundamentals of Inhomogeneous Fluids}, edited by D.~Henderson (Marcel Dekker Inc., NY, 1992), p.85. 

\bibitem{Brezin2010}
E.~Br\'ezin, S. Franz, and G. Parisi, 
Phys. Rev. B {\bf 82}, 144427 (2010).

\bibitem{NanoWodka}
N.~Severin, J.~Gienger, V.~Scenev, P.~Lange, I. M.~Sokolov, and J. P.~Rabe, 
Nano Lett. {\bf 15}, 1171 (2015).

\bibitem{Rayermann}
S. P.~Rayermann, G. E.~Rayermann, C. E.~Cornell, A. J.~Merz, and S. L.~Keller,
Biophys. J. {\bf 113}, 2425 (2017).


\bibitem{Shishulin}
A. V.~Shishulin, and V. B.~Fedoseev, 
J. Mol. Liq. {\bf 278}, 363 (2019). 


\bibitem{Puibasset}
J.~Puibasset, 
J. Chem. Phys. {\bf 126}, 184701 (2007).


\bibitem{Pressing}
J.~Pressing, and J. E. Mayer, 
J. Chem. Phys., {\bf 59}(5), 2711 (1973).


\bibitem{BrezinFeng}
E.~Br\'ezin, and S.~Feng, 
Phys. Rev. B {\bf 29}, 472 (1984).


\bibitem{Jug}
G.~Jug, and D.~Jasnow, 
Phys. Rev. B {\bf 31}, 1610 (1985).


\bibitem{Moldover}
H.~Chaar, M.~Moldover and J.~Schmidt, 
J. Chem. Phys. {\bf 85}, 418 (1986).

\bibitem{Mon} K. K.~Mon, Phys. Rev. Lett. {\bf 60}, 2749 (1988).


\bibitem{Muenster}
G.~M\"unster, 
Nucl. Phys. B {\bf 340}, 559 (1990).


\bibitem{Brackbill}
J. U.~Brackbill, D. B.~Kothe, and C.~Zemach, 
J. Comp. Phys. {\bf 100}, 335 (1992). 


\bibitem{Mecke}
M.~Mecke, J.~Winkelmann, and J.~Fischer, 
J. Chem. Phys. {\bf 107}, 9264 (1997). 

\bibitem{Potoff}
J. J.~Potoff, and A. Z.~Panagiotopoulos, 
J. Chem. Phys. {\bf 112}, 6411 (2000).



\bibitem{Brilliantov2002} N. V.~Brilliantov, J. M.~Rubi, 
arXiv: 0201340 [cond-mat.soft].

\bibitem{Caillol2010}
V.~Russier, and J. M.~Caillol,
Cond. Matt. Phys. {\bf 13}:2, 23602:1 (2010).


\bibitem{Ghoufi}
A.~Ghoufi, and P.~Malfreyt
J. Chem. Phys. {\bf 146}, 084703 (2017).


\bibitem{Goujon}
F.~Goujon, A.~Ghoufi, P.~Malfreyt, 
Chem. Phys. Lett. {\bf 694}, 60 (2018).


\bibitem{Popinet}
S.~Popinet, Annual Rev. of Fluid Mech. {\bf 50(1)}, 49 (2018).


\bibitem{Hernandez-Munoz}
J.~Hern\'andez-Mu\~noz, P.~Tarazona, R.~Ram\'irez, C. P.~Herrero, and E.~Chac\'on, 
Phys. Rev. B {\bf 100}, 195424 (2019).


\bibitem{Eltsov}
V. B.~Eltsov, A.~Gordeev, and M.~Krusius,
Phys. Rev. B {\bf 99}, 054104 (2019).



\bibitem{Lo}
H. Y.~Lo, Y.~Liu, S. Y.~Mak, Zh.~Xu, Y.~Chao, K. J.~Li, H. Ch.~Shum, and L.~Xu, 
Phys. Rev. Lett. {\bf 123}, 134501 (2019).


\bibitem{Kaurin}
D.~Kaurin, and M.~Arroyo, 
Phys. Rev. Lett. {\bf 123}, 228102 (2019).


\bibitem{Zinn-Justin}
J.~Zinn-Justin, 
{\it Quantum Field Theory and Critical Phenomena} (Clarendon Press, Oxford, UK, 2002).

\bibitem{Brezin}
E.~Br\'ezin, {\it Introduction to Statistical Field Theory} (Cambridge University Press, Cambridge, UK, 2010).

\bibitem{Parola}
A.~Parola, and L.~Reatto,
Adv. in Phys. {\bf 44}:3, 211 (1995).

\bibitem{Pelissetto}
A.~Pelissetto, and E.~Vicari,
Phys. Rep. {\bf 368}, 549 (2002).


\bibitem{BrezinZinn}
E.~Br\'ezin, J. C.~Le Guillou, J.~Zinn-Justin, in {\it Phase Transitions and Critical Phenomena}, edited by C.~Domb, and M. S.~Green (Academic Press, London, UK, 1976),  {\bf 6}, p. 127.


\bibitem{WilsonKogut}
K. G.~Wilson, and J.~Kogut, 
Phys. Rep. {\bf 12c}, 75 (1974).


\bibitem{Hubbard}
J.~Hubbard, and P.~Schofield, 
Phys. Lett., A {\bf 40}, 245 (1972).


\bibitem{WCA}
H. C.~Andersen, D.~Chandler, and J. D.~Weeks,
J. Chem. Phys. {\bf 56}, 3812 (1972).


\bibitem{Gubbins}
C. G.~Gray, and K. E.~Gubbins, 
{\it Theory of molecular fluids} (Clarendon Press, Oxford, UK, 1984).

\bibitem{Trokhym}
A.~Trokhymchuk, R.~Melnyk, M.~Holovko, J.~Nezbeda, 
J. Mol. Liq., {\bf 228}, 194 (2017).


\bibitem{Brilliantov1998}
N. V.~Brilliantov, 
Phys. Rev. E {\bf 58}, 2628 (1998).


\bibitem{Holcomb}
C. D.~Holcomb, P.~Clancy, and J. A.~Zollweg, Mol. Phys. {\bf 78}, 437 (1993).


\bibitem{Amit}
D. J.~Amit, 
{\it Field Theory: Renormalization Group and
Critical Phenomena} (McGraw-Hill, NY, 1978).


\bibitem{Kubo}
R.~Kubo,
J. Phys. Soc. Japan {\bf 17}, 1100 (1962).


\bibitem{Meeron}
E.~Meeron,
J. Chem. Phys. {\bf27}, 1238 (1957).

\end{thebibliography}
\end{document}